\documentclass[preprint,pre]{revtex4}
\usepackage[T1]{fontenc}
\usepackage[latin1]{inputenc}
\usepackage[dvips]{graphicx}
\usepackage{float}
\usepackage{amssymb}
\usepackage{amsmath}

\newcommand{\lav}{\left\langle}
\newcommand{\rav}{\right\rangle}

\restylefloat{figure}

\begin{document}
\title{Molecular Traffic Control in a 3D network of single file channels and fast reactivity}
\author{Andreas Brzank\textsuperscript{1,3}}
\email{a.brzank@fz-juelich.de}
\author{Sungchul Kwon\textsuperscript{2}}
\author{Gunter Schütz\textsuperscript{1}}
\affiliation{
\textsuperscript{1}Institut f\"ur Festk\"orperforschung, Forschungszentrum J\"ulich, 52425 J\"ulich, Germany\\
\textsuperscript{2}Department of Physics, Kyung Hee university, Seoul 130-701, Korea\\
\textsuperscript{3}Fakult\"at für Physik und Geowissenschaften, Universit\"at Leipzig, Abteilung Grenzfl\"achenphysik,
Linnestrasse 5, D-04103 Leipzig, Germany}
\date{\today}

\begin{abstract}
{

We study the conditions for reactivity enhancement of catalytic processes in porous solids
by use of molecular traffic control (MTC) as a function of grain size. We extend a recently 
introduced two dimensional model system to three dimensions. With dynamic Monte-Carlo simulations 
and analytical solution of the associated Master equation we obtain a quantitative description of 
the MTC effect in the limit of fast reactivity. The efficiency ratio (compared with a topologically
and structurally similar reference system without MTC) is inversely proportional to the grain diameter.
}
\end{abstract}
\maketitle

\section{Introduction}
Zeolites are used for catalytic processes in a variety of applications, e.g.
cracking of large hydrocarbon molecules. In a number of zeolites diffusive
transport occurs along quasi-one-dimensional channels which do not allow guest
molecules to pass each other \cite{Baer01}. Due to mutual blockage of reactands
$A$ and product molecules $B$ under such {\it single-file conditions}
\cite{Karg92} the effective reactivity of a catalytic process $A\to B$ -- determined
by the residence time of molecules in the zeolite -- may be considerably reduced
as compared to the reactivity in the absence of single-file behaviour.
It has been suggested that the single-file effect may be circumvented by the so far
controversial concept of molecular traffic control (MTC) \cite{Dero80,Dero94}.
This notion rests on the assumption that reactands and product molecules resp.
may prefer spatially separated diffusion pathways and thus avoid mutual
suppression of self-diffusion inside the grain channels.

The necessary (but not sufficient) requirement for the MTC effect, a channel
selectivity of two different species of molecules, has been verified by
means of molecular dynamic (MD) simulations of two-component mixtures
in the zeolite ZSM-5 \cite{Snur97} and relaxation simulations of a
mixture of differently sized molecules (Xe and SF$_6$) in a bimodal
structure possessing dual-sized pores (Boggsite with 10-ring and 12-ring pores)
\cite{Clar00}. Also equilibrium Monte-Carlo simulations demonstrate that the
residence probability in different areas of the intracrystalline pore space
may be notably different for the two components of a binary mixture
\cite{Clar99} and thus provide further support for the notion
of channel selectivity in suitable bimodal channel topologies.

\begin{figure}
\centerline{
\includegraphics[width=6cm]{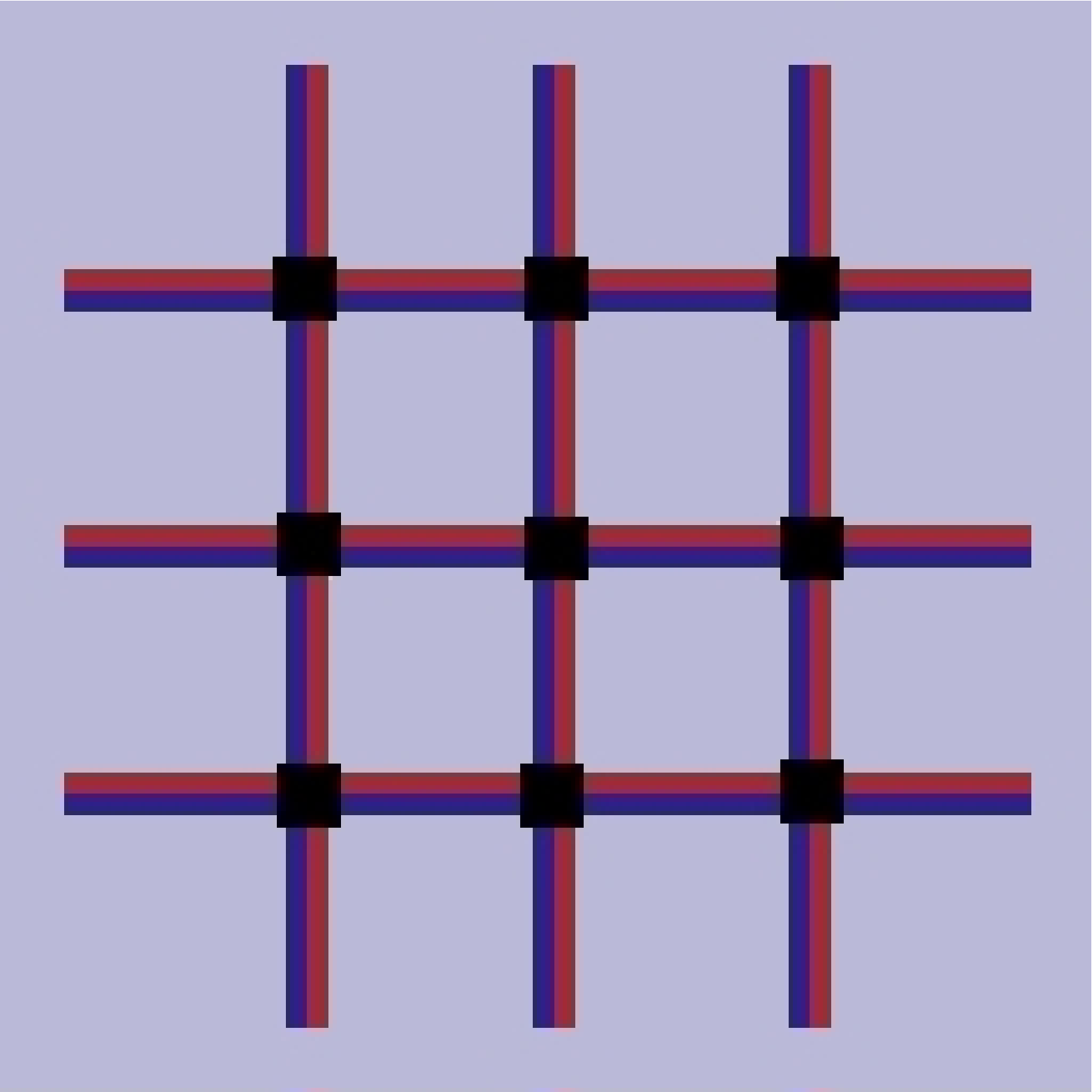}
\includegraphics[width=6cm]{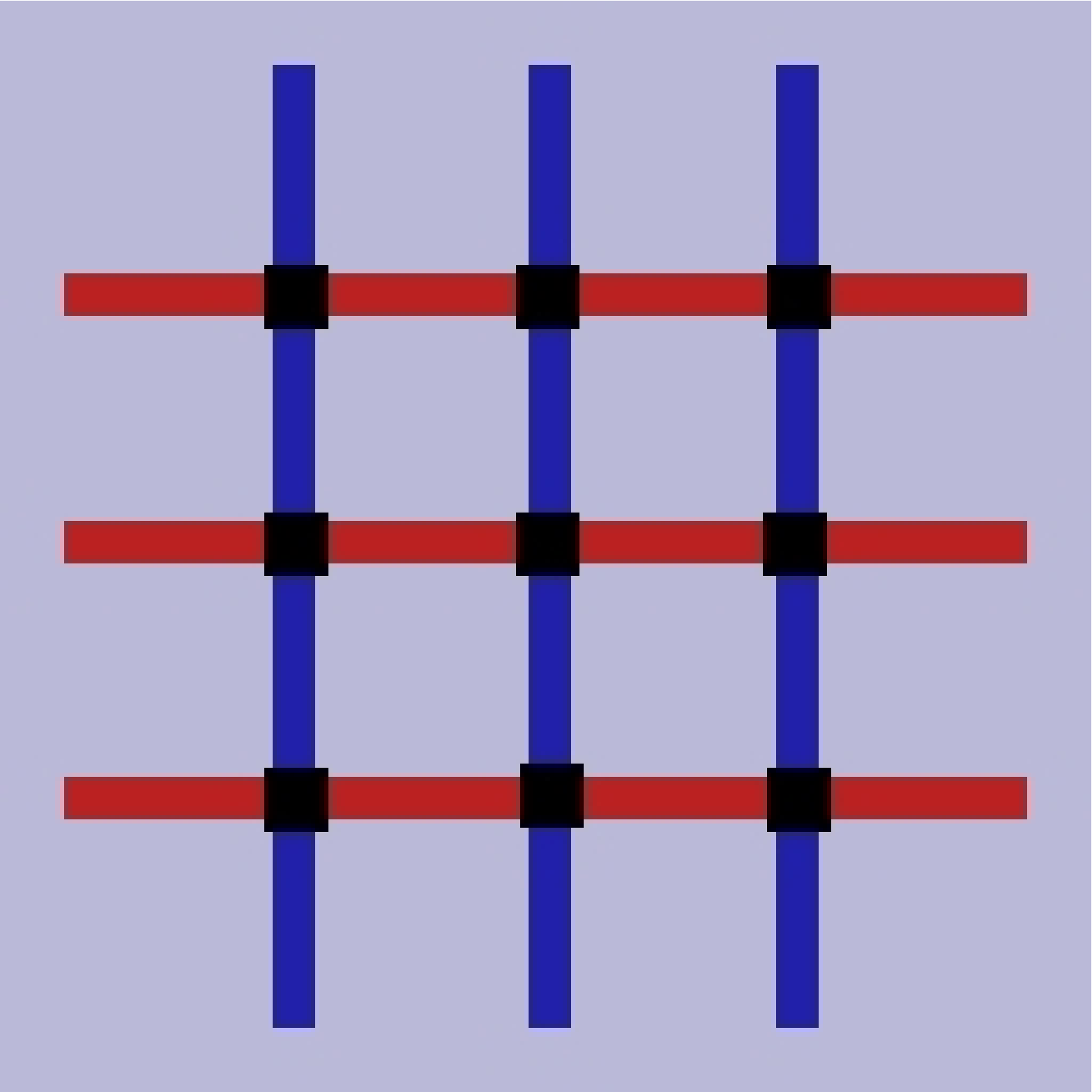}
}
\caption{REF system (left) with $N=3$ channels and MTC system
(right) of the the same size. In contrast to the REF case, where we allow both
types of particles ($A$ and $B$ particles) to enter any channel, in the MTC
system $A$ particles are carried through the vertical $\alpha$
channels whereas the $B$ particles diffuse along the horizontal $\beta$
channels. Black squares indicate catalytic sites where a catalytic
transformation $A\to B$ is allowed.}
\label{systemPics}
\end{figure}

Whether a MTC effect leading to reactivity enhancement actually takes place was
addressed by a series of dynamic Monte Carlo simulations (DMCS) of a stochastic
model system with a network of perpendicular sets of bimodal intersecting channels
and with catalytic sites located
at the intersecting pores (NBK model) \cite{Neug00,Karg00,Karg01}. The authors
of these studies found numerically the occurrence of the MTC effect by comparing
the outflow of reaction products in the MTC system with the outflow from
a reference system with equal internal and external system parameters, but no
channel selectivity (Fig. \ref{systemPics}). The dependency of the MTC effect as a
function of the system size has been investigated in \cite{Brau03}.
The MTC effect is favored by a small number of channels and occurs only for
long channels between intersections, which by themselves lead to a very low
absolute outflow compared to a similar system with shorter channels. A recent
analytical treatment of the master equation for this stochastic many-particle
model revealed the origin of this effect at high reactivities \cite{Brza04}.
It results from an interplay of the long residence time of guest molecules
under single-file conditions with a saturation effect that leads to a depletion
of the bulk of the crystallite. Thus the MTC effect is firmly established, but the
question of its relevance for applications remains open.

Here we address this question by an analytical study of the MTC system in three dimensions as a
function of grain size. This may be of interest as since the first successful
synthesis of mesoporous MCM-41 nanoparticles \cite{Beck92}, there has been
intense research activity in the design and synthesis of structured mesoporous
solids with a controlled pore size. In particular, synthesis of bimodal nanostructures
with independently controlled small and large mesopore sizes has become
feasible \cite{Sun03}.

\section{NBK Model}
\begin{figure}
\centerline{\includegraphics[width=7cm]{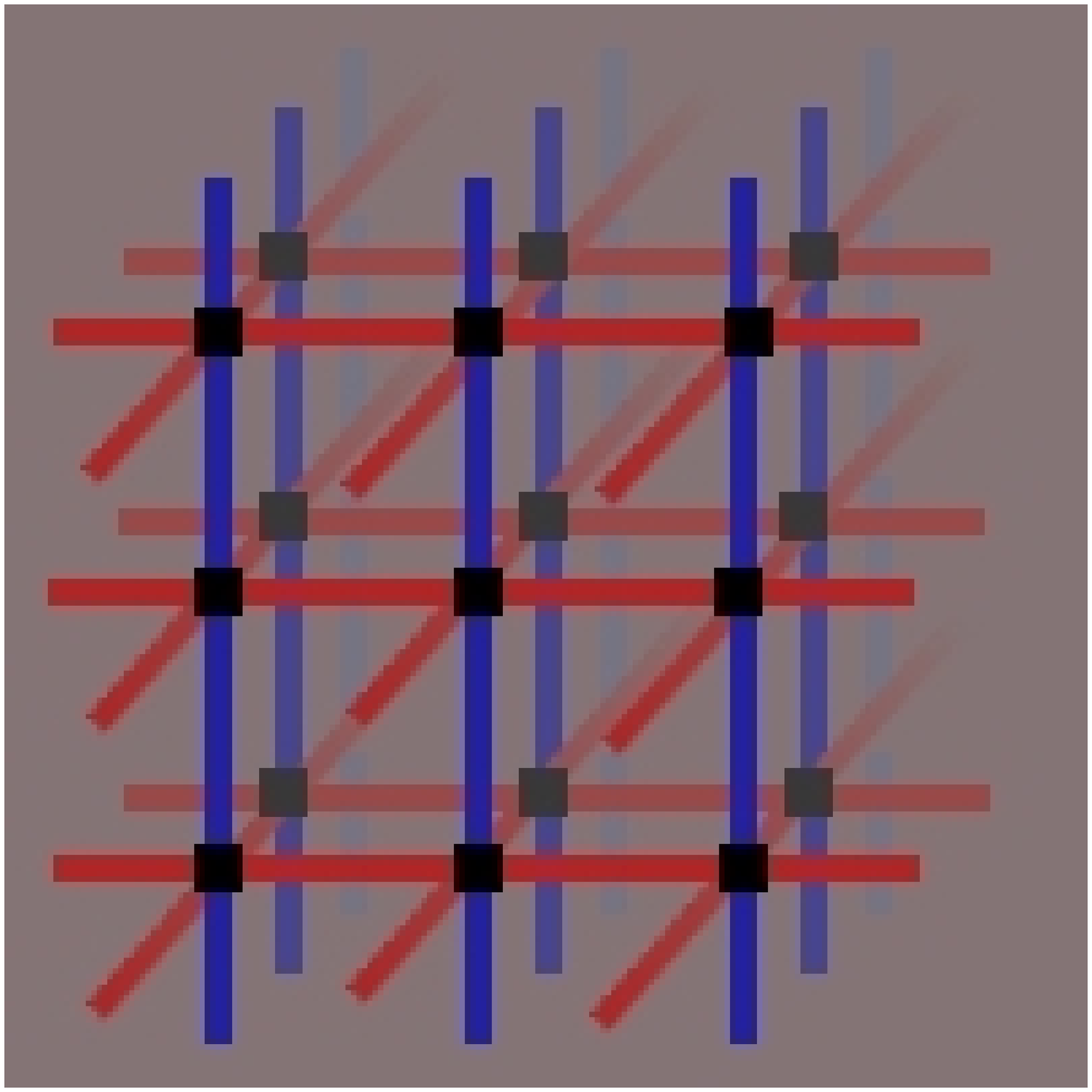}
\includegraphics[width=7cm]{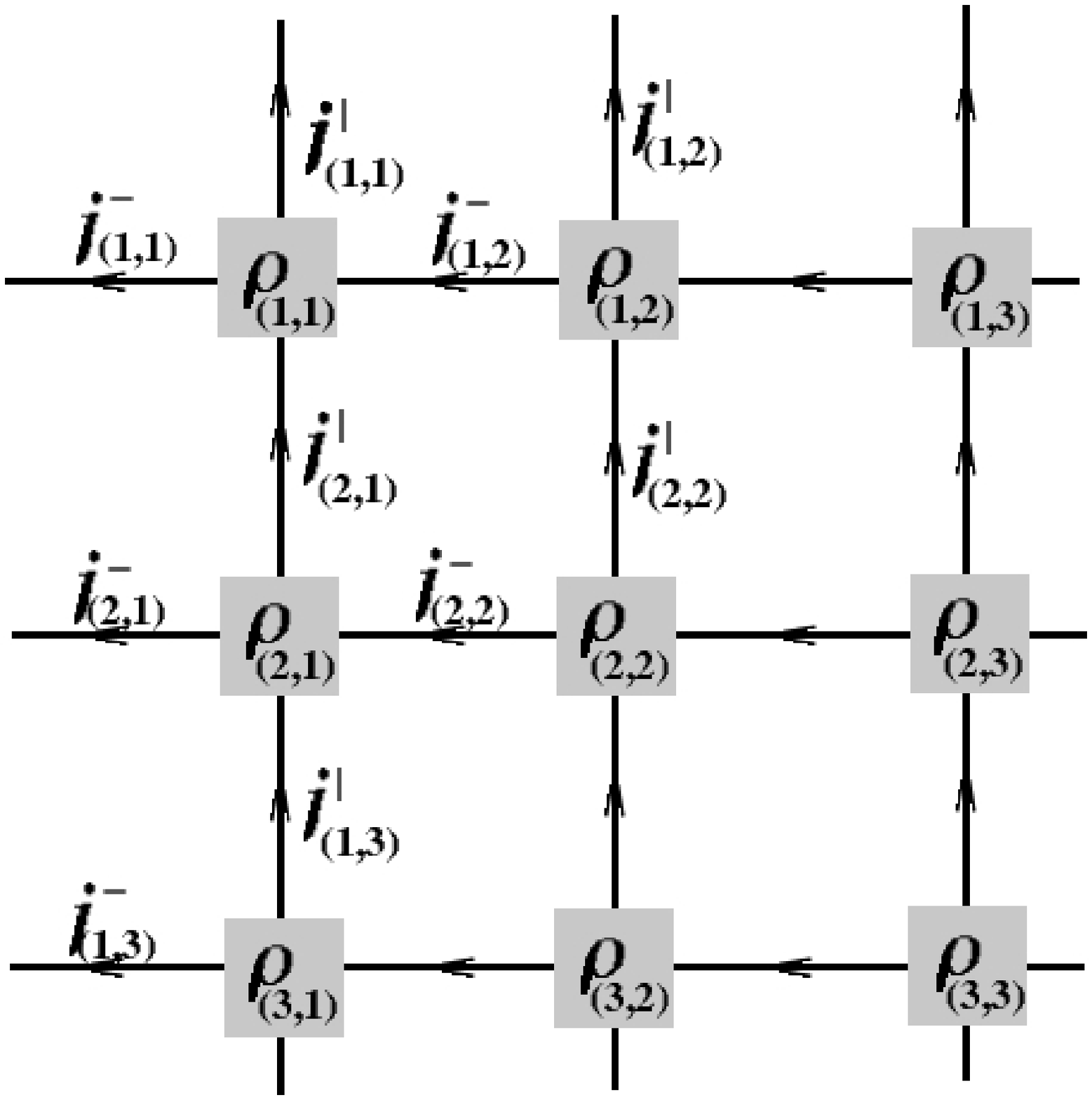}}
\caption{(left) 3D MTC system. (right) part of the first $y/z$ plane of $\beta$ channels, view from top.}
\label{mtc3D}
\end{figure}

Similar to \cite{Brau03,Brza04,Neug00} we consider the NBK lattice model as an array of $N\times N\times N$
channels (Fig. \ref{mtc3D} left) which is a measure of the grain size of the crystallite.
Each channel has $L$ sites between the intersection points where the irreversible catalytic process
$A\to B$ takes place. We assume the boundary channels of the grain to be connected to the surrounding gas
phase, modelled by reservoirs of constant densities such that the entrances of the respective channels
(extra reservoir sites) have a fixed $A$ particle density $\rho$. We assume the reaction products $B$
which leave the crystallite to be removed immediately from the gas phase such that the density of $B$
particles in the reservoir is always 0. Short-range interaction between
particles inside the narrow pores is described by an idealized hard core
repulsion which forbids double-occupancy of lattice sites.

The underlying dynamics are stochastic. We work in a continuous time description
where the transition probabilities become transition rates and no multiple
transitions occur at the same infinitesimal time unit. Each elementary transition
between microscopic configurations of the system takes place randomly with an
exponential waiting-time distribution. Diffusion is modelled by jump processes
between neighbouring lattice sites. $D$ is the elementary (attempt) rate of
hopping and is assumed to be the same for both species $A,B$ of particles.
In the absence of other particles $D$ is the self-diffusion coefficient for the 
mean-square displacement along a channel. If a neighboring site is occupied by a 
particle then a hopping attempt is rejected (single-file effect). The dynamics 
inside a channel are thus given by the symmetric exclusion process
\cite{Spit70,Spoh83,vanB83,Schu94} which is well-studied in the probabilistic
\cite{Ligg99} and statistical mechanics literature \cite{Schu01}. The
self-diffusion along a channel is anomalous, the effective diffusion rate between
intersection points decays asymptotically as $1/L$, see \cite{vanB83} and references therein.

At the intersections the reaction $A\to B$ occurs with a reaction rate $c$.
This reaction rate influences, but is distinct from, the effective grain
reactivity which is largely determined by the residence time of guest
molecules inside the grain which under single-file conditions grows in the
reference system with the third power of the channel length $L$
\cite{Rode99,vanB83}. At the boundary sites particles jump into the reservoir with
a rate $D(1-\rho_A-\rho_B)$ in the general case. Correspondingly
particles are injected into the grain with rates $D\rho_{A,B}$ respectively.
As discussed above here we consider only $\rho_A=\rho$, $\rho_B=0$.

For the REF system $A$ and $B$ particles are allowed to enter and leave both types 
of channels, the blue ($\alpha$) and red ($\beta$) ones (Fig. \ref{mtc3D} left).
In case of MTC $A$($B$) particles will enter $\alpha$($\beta$)-channels only,
mimicking complete channel selectivity. Therefore all channel segments carry
only one type of particles in the MTC case. For the boundary channels
complete selectivity implies that $\alpha$-channels are effectively
described by connection with an $A$-reservoir of density $\rho_A=\rho$
($B$-particles do not block the boundary sites of $\alpha$-channels)
and $\beta$-channels  are effectively described by connection with a
$B$-reservoir of density $\rho_B=0$, respectively.
($A$-particles do not block the boundary sites of $\beta$-channels.)
This stochastic dynamics, which is a Markov process, fully defines the NBK
model.

In both cases, MTC and REF system, the external concentration gradient
between $A$ and $B$ reservoir densities induces a particle current inside the
grain which drives the system into a stationary nonequilibrium state.
For this reason there is no Gibbs measure and equilibrium
Monte-Carlo algorithms cannot be applied for determining steady state
properties. Instead we use dynamic Monte-Carlo simulation (DMCS)
with random sequential update. This ensures that the simulation algorithm yields
the correct stationary distribution of the model.

\section{MTC in 3D with large reactivity}
Anticipating concentration gradients between intersection points we expect due
to the exclusion dynamics linear density profiles within the channel
segments \cite{Spoh83,Schu01,Brza04}, the slope and hence the current
being inversely proportional to the number of lattice sites $L$.
The total output current $j$ of $B$ particles, defined as the number of
$B$-particles leaving the grain per time unit in the stationary state, is the main
quantity of interest. It determines the effective reactivity of the grain.

We are particularly interested in studying the system in its maximal
current state for given reactivity $c$ and size constants $N$, $L$, which are
intrinsic material properties of a given grain. The $A$ particle reservoir
density $\rho$, determined by the density in the gas phase, can be tuned in
a possible experimental setting. Let us therefore denote the reservoir density
which maximizes the output current by $\rho^*$ and the maximal current by $j^*$.
For MTC systems as defined above we always expect $\rho_{MTC}^*=1$,
since the highly charged entrances of $\alpha$-channels do not block the exit
of $B$-particles and hence do not prevent them from
leaving the system. 

In order to measure the efficiency of a MTC system over the associated REF system
we define the efficiency ratio
\begin{align}
R(c,N,L)=\frac{j_{MTC}^*}{j_{REF}^*}
\end{align}
which is a function of the system size $N$, $L$ and reactivity $c$.

\begin{figure}
\centerline{\includegraphics[width=13cm,angle=0]{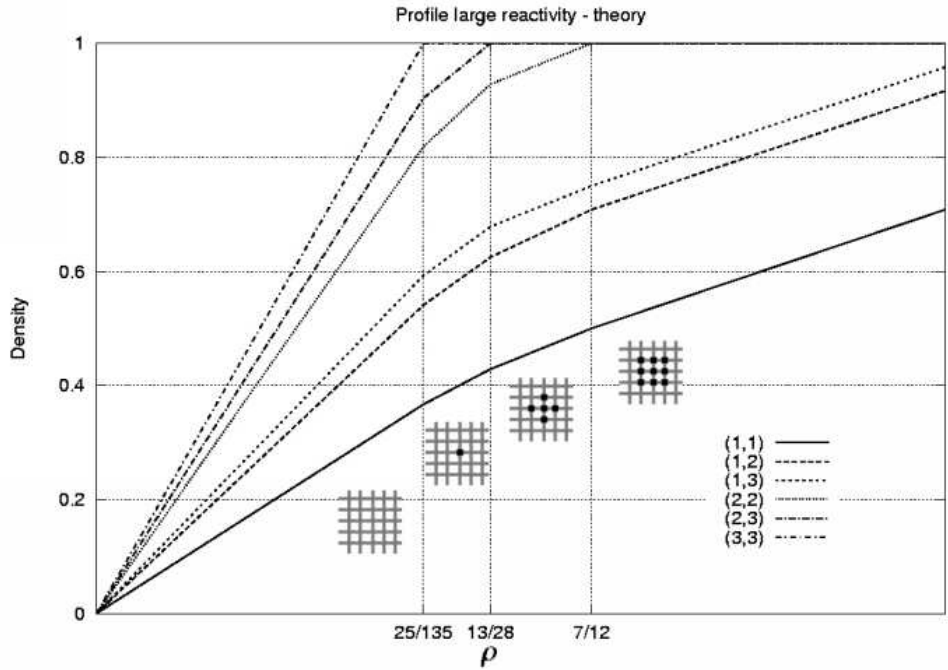}}
\caption{Theoretical intersection densities for a lattice of $N=5$. The regimes of saturating intersections
are indicated.}
\label{3dprofile}
\end{figure}

\begin{figure}
\centerline{\includegraphics[width=15cm,angle=0]{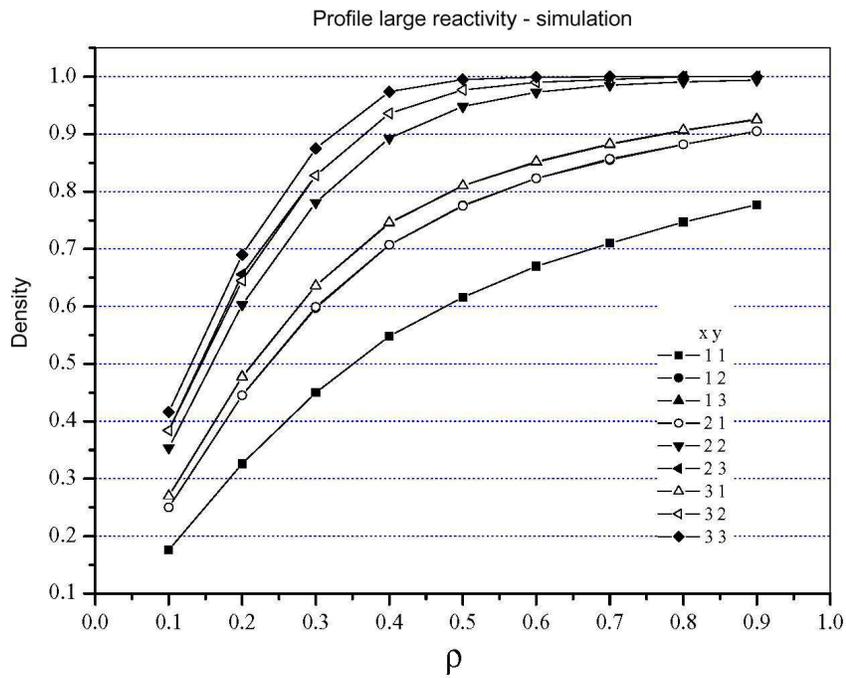}}
\caption{Simulated intersection densities for a lattice of $N=5$ and $L=10$.}
\label{3dsimfinal}
\end{figure}

Let us now discuss the fast reactivity case. The penetration of $A$ particles is controlled
by $c$ and in the limit of $c\to\infty$ $A$ particles entering the system will be
converted as soon as they reach the first intersection. Therefore, only the first and the last
plane of $\beta$-channels contribute to the $B$ particle output. The profile as well as the $B$ particle
output is fully determined by the intersection densities $\rho_{(x,y)}$. Fig. \ref{mtc3D} (right)
shows the top view of the first plane indicating our notation of the densities and currents. 
$\alpha$-channels point into the plane and are not displayed. The current of an
$\alpha$-channel segment connecting the reservoir with intersection $\rho_{(x,y)}$ is denoted
by $j^A_{(x,y)}$. For an analysis of the Master equation for this process we neglegt correlations 
between the occupancy of a catalytic site and its five neighbours. This mean field 
approximation is motivated by exact results for the correlations in the stationary state 
of the symmetric exclusion process from which it is known \cite{Schu01} that nearest neighbour correlations
in the vicinity of the boundary of a system of size $L$ are of order $1/L^2$. Within mean field 
we replace joint probabilities $\lav xy\rav$ by the product $\lav x\rav\lav y\rav$.
For the stationary state we identify the following currents,
\begin{align}
\label{currents}
j^A_{(x,y)}&=D\frac{\rho\left(\rho_{(x,y)}-1\right)}{L\left(\rho_{(x,y)}-1\right)-1}\\
j^-_{(x,y)}&=D\frac{\rho_{(x+1,y)}-\rho_{(x,y)}}{L+1}\\
j^|_{(x,y)}&=D\frac{\rho_{(x,y+1)}-\rho_{(x,y)}}{L+1}
\end{align}
where $\rho$ is the reservoir density of $A$ particles. Using conservation of currents
\begin{align}
\label{conservation}
j^A_{(x,y)}+j^-_{(x+1,y)}+j^|_{(x,y+1)}=j^-_{(x,y)}+j^|_{(x,y)}
\end{align}
leads to a set of quadratic equations for the intersection densities $\rho_{(x,y)}$.
For large $L$ one of the two solutions reduces to a discrete Poisson equation.
\begin{align}
\label{recursion3D}
{\rho}_{(x,y)}=\frac{1}{4}\left({\rho}_{(x+1,y)}+{\rho}_{(x-1,y)}+{\rho}_{(x,y+1)}+{\rho}_{(x,y-1)}+{\rho} \right).
\end{align}
The other solution $\rho_{(x,y)}=1$ demonstrates the fact that due to exclusion only one particle per site is allowed.
\eqref{recursion3D} can be solved by use of the discrete sine transform
$\tilde{\rho}_{(q,p)}=\sum_{x=1}^N\sum_{y=1}^N \rho_{(x,y)}\sin\frac{q\pi x}{N+1}\sin\frac{p\pi y}{N+1}$.
We express \eqref{recursion3D} in terms of the  transformed density $\tilde{\rho}_{(q,p)}$. Taking
into account the boundary conditions $\rho_{0,y}=\rho_{x,0}=\rho_{x,N+1}=\rho_{N+1,y}=0$ with $0\le x,y \le N+1$ we find
\begin{align}
\label{diffeqtrns}
\tilde{\rho}_{(q,p)}=\frac{2\rho}
{\cos\frac{q\pi}{N+1}+\cos\frac{p\pi}{N+1}-2}\sum_{n=1}^N\sum_{m=1}^N\sin\frac{q\pi n}{N+1}\sin\frac{p\pi m}{N+1}.
\end{align}
The non zero contributions of the double sum can be expressed as a product of two Cotangens.
Transforming back finally yields
\begin{align}
\label{REFsolution}
{\rho}_{(x,y)}&\equiv\rho M_{(x,y)}=-\frac{2\rho}{(N+1)^2}\sum_{n=1}^N\sum_{m=1}^N\frac{B_{(n,m)}}{(\cos\frac{n\pi}{N+1}+\cos\frac{m\pi}{N+1}-2)}\sin\frac{n\pi x}{N+1}sin\frac{n\pi y}{N+1}\\
B_{(n,m)}&=
\begin{cases}
0 & \text{if $n$ or $m$ even} \\
\cot\frac{m\pi}{2(N+1)}\cot\frac{n\pi}{2(N+1)} & \text{else}
\end{cases}
\end{align}
However, similar to MTC in 2D \cite{Brza04}, increasing the boundary density $\rho$ eventually leads intersections to saturate starting from the most inner one. Thus, \eqref{REFsolution} is true only for small $\rho$. To be more
precise \eqref{REFsolution} is a solution for $0\le\rho\le\frac{1}{M_{([N+1]/2,[N+1]/2)}}$. A further increase of
$\rho$ eventually leads other intersection to saturate. Fig. \ref{3dprofile} shows the theoretical densities (system with
$N=5$-channels) as a function of $\rho$. We identify different regimes of saturating intersections.
We observe a grouping $\{(1,1)\}, \{(1,2),(1,3)\}$ and $\{(2,2),(2,3),(3,3)\}$ which is also supported by simulations (Fig. \ref{3dsimfinal} same system size and slightly different model rates), where only
intersections of the last group saturate. The smoothness of the simulated curves is due to
finite-size effects which are not captured within mean field theory.


The situation is similar to the case in two dimensions. The main constrained of the
MTC system becomes apparent as we consider large reactivities. The output is
mainly determined by two planes of $\beta$-channels. $B$ particles leave the system through $8N$
channel segments. In the reference system the output is proportional to $N^2$. Therefore a
MTC effect is expected only for systems with small number of channels. Increasing the distance
$L$ would in principle favor MTC systems but reduces the absolute output. Simulations also show
no evidence for a qualitative different picture compared to MTC in 2D.

For fixed moderate $c$ this extreme situation is not realized, but nevertheless
with increasing $N$ one expects that the bulk gets increasingly depleted,
since in each layer a fraction of $A$ particles is converted into $B$
particles. Thus fo lage $N$ the total $A$-density in each layer may be described 
in a continuum desciption by the form
\begin{align}
\frac{d}{dx} N_A(x) = - \gamma N_A(x)
\end{align}
for the number $N_A(x)$ of $A$-particles in layer $x$ predicts an
exponential decrease of the $A$ density,
leaving only an active boundary layer of finite thickness
\begin{align}
\xi = 1/\gamma \propto 1/c
\end{align}
at the top and
bottom respectively of the (in our simulation three-dimensional) grain.
Hence, as a function of $N$, $j^\ast_{MTC}$ saturates at some constant
\begin{align}
\lim_{N \to\infty} j^\ast_{MTC}(c,N,L)= C^\ast_{MTC}(c,L).
\end{align}

On the other hand, in the REF system the output current scales linearly with
increasing $N$ for all, even large, $c$. This is because
even though again the bulk depletes with increasing $N$ the active boundary
layer is a surface scaling linearly with $N$. Thus
\begin{align}
\lim_{N \to\infty} j^\ast_{REF}(c,N,L)= N C^\ast_{REF}(c,L)
\end{align}
Hence
\begin{align}
R(c,N,L) \propto 1/N
\end{align}
and the MTC effect vanishes at some $N$ for fixed reactivity $c$
and channel length $L$.

\section{Conclusion}
We have generalized the two-dimensional NBK-model for molecular traffic
control to three space dimensions. We have focussed our attention
on the case of short intersecting channels between catalytic centers which is
the relevant setting from an applied perspective. Inspired by large deviation theory
for nonequilibrium steady states and using exact results for the SEP we reduce the
large number of freedom to a system of equations for the
effective densities at the reaction sites. Our analytical treatment for
large reactivities yields stationary density profiles which are suppoted by
simulations. As in two dimensions \cite{Brza04} a sequence of
surface induced saturations of the reaction sites inside the crystallite sets in.
This leads to nonanalytical changes of the output of product molecules as a
function of the input rate of reactands.

For moderate reactivities we obtain
an exponentially decreasing loading as one probes the system further
away from the boundary where reactands are adsorbed.
The localization length is inversely proportional to the
local reaction rate. As a consequence, the effective reactivity of
crystallites with a diameter larger than the localization length does not
scale with the surface area of the crystallite, but only with the diameter.
Therefore, for moderate and high local reactivities and grain sizes currently
used in industrial processes one expects no reaction enhancement through the
MTC effect. However, as proposed in \cite{Brza05}, nano-sized
grains do exhibit this phenomenon and are thus potential candidates
for exploiting MTC.


\begin{thebibliography}{10}

\bibitem{Baer01}
Ch. Baerlocher, W. M. Meier and D. H Olson, Atlas of Zeolite Structure Types, Elsevier: London 2001.

\bibitem{Karg92}
J. K\"arger and D.M. Ruthven, Diffusion in Zeolites and Other Microporous Solids, Wiley: New York 1992.

\bibitem{Dero80}
E.G. Derouane and Z. Gabelica, J. Catal. 65 (1980) 486.

\bibitem{Dero94}
E.G. Derouane, Appl. Catal. A, N2 (1994) 115.

\bibitem{Snur97}
R. Q. Snurr and J. Kärger, Phys. Chem. B 101 (1997) 6469.

\bibitem{Clar00}
L. A. Clark,G. T. Ye, R. Q. Snurr, Phys. Rev. Lett. 84 (2000) 2893.

\bibitem{Clar99}
L. A. Clark,G. T. Ye,A. Gupta,L. L. Hall,R. Q. Snurr, J. Chem. Phys. 111 (1999) 1209.

\bibitem{Heuc97}
M. Heuchel, R.Q. Snurr, E. Buss, Langmuir 13 (1997) 6249.

\bibitem{Neug00}
N. Neugebauer, P. Br\"auer, J. K\"arger, J. Catal. 194 (2000) 1.

\bibitem{Karg00}
J. K\"arger, P. Br\"auer, H. Pfeifer, Z. Phys. Chem. 104 (2000) 1707.


\bibitem{Karg01}
J. Kärger, P. Bräuer, A. Neugebauer, Europhys. Lett. 53 (2001) 8.

\bibitem{Brau03}
P. Bräuer, A. Brzank,J. Kärger, J. Phys. Chem. B 107 (2003) 1821.

\bibitem{Brza04}
A. Brzank, G. M. Schütz, P. Bräuer, and J. Kärger, Phys. Rev. E 69 (2004) 031102.

\bibitem{Beck92}
J.S. Beck, J.C. Vartulli, W.J. Roth, M.E. Leonowice, C.T. Kresge,
K.D. Schmitt, C. T-W. Chu, D.H. Olsen, E.W. Sheppard, S.B. McCullen,
J.B. Higgins, J.L. Schlenker, J. Am. Chem. Soc. 114 (1992) 10834.

\bibitem{Sun03}
J.H. Sun, Z. Shan, Th. Maschmeyer, M.-O. Coppens, Langmuir 19 (2003) 8395.

\bibitem{Spit70}
F. Spitzer, Adv. Math. 5 (1970) 246.

\bibitem{Spoh83}
H. Spohn, J. Phys. A 16 (1983) 4275.

\bibitem{vanB83}
H. van Beijeren, K.W. Kehr, R. Kutner, Phys. Rev. B 28 (1983) 5711.

\bibitem{Schu94}
G. Sch\"utz, S. Sandow, Phys. Rev. E 49 (1994) 2726.

\bibitem{Ligg99}
T.M. Liggett: {\it Stochastic Interacting Systems: Contact, Voter and
Exclusion Processes} (Springer, Berlin, 1999).

\bibitem{Schu01}
G.M. Sch\"utz, in: {\it Phase Transitions and
Critical Phenomena}, C. Domb and J. Lebowitz (eds.),
(Academic, London, 2001).

\bibitem{Rode99}
C. R\"odenbeck, J. K\"arger, J. Chem. Phys. 110 (1999) 3970.

\bibitem{Brza05}
A. Brzank, G.M. Sch\"utz, Appl. Catalysis A in press
\end{thebibliography}
\end{document}